\documentclass[prl,twocolumn]{revtex4-1}
\usepackage{bm}
\usepackage{epsfig}
\usepackage{amsmath}
\usepackage{latexsym}
\usepackage{slashed} 

\newcommand{\be}{\begin{equation}}
\newcommand{\ee}{\end{equation}}
\newcommand{\bea}{\begin{eqnarray}}
\newcommand{\eea}{\end{eqnarray}}
\newcommand{\bma}{\begin{matrix}}
\newcommand{\ema}{\end{matrix}}
\newcommand{\nn}{\nonumber}
\newcommand{\bml}{\begin{mathletters}}
\newcommand{\eml}{\end{mathletters}}
\newcommand{\bes}{\begin{subequations}}
\newcommand{\ees}{\end{subequations}}

\newcommand{\bi}{\begin{itemize}}
\newcommand{\ei}{\end{itemize}}

\begin{document}

\title{Dynamical Electroweak Symmetry Breaking with a Heavy Fourth Generation}
\author{P.Q. Hung}
\email[]{pqh@virginia.edu} 
\author{Chi Xiong}
\email[]{xiong@virginia.edu}
\affiliation{Dept. of Physics, University of Virginia, \\
382 McCormick Road, P. O. Box 400714, Charlottesville, Virginia 22904-4714,
USA}

\date{\today}
\begin{abstract}
A heavy fourth generation with a mass of the order of 400 GeV or more could trigger dynamical electroweak symmetry breaking by forming condensates through the exchange of a fundamental Higgs scalar doublet.  The dynamics leading to these condensates is studied within the framework of the Schwinger-Dyson equation. This scenario leads to the presence of {\em three} (two composite and one fundamental) Higgs doublets, with interesting phenomenological implications.  In addition, this dynamical phenomenon occurs in the vicinity of the energy scale where the restoration of scale symmetry might happen.
\end{abstract}

\maketitle
\section{Introduction}

The nature of the electroweak symmetry breaking is one of the- if not the most- pressing issue in particle physics. The Large Hadron Collider (LHC) was built largely out of the desire to discover this ``holy grail" of high energy physics: The Higgs particle. In its simplest form, the Higgs boson of the Standard Model (SM) is an elementary particle whose mass is not predicted by the SM. Its discovery as well as the subsequent studies of its property are without any doubt the most important aspects of particle physics for at least the next decade. 

In the meantime, there are a number of often-asked questions that one would like to address and which hopefully could be answered experimentally in a not-too-distant future. The first question is the very concept of spontaneous symmetry breaking (SSB) as applied to the SM. Although  a number of arguments point to the possible correctness of SSB,  it is not implausible that there is a theory that might mimic many of its results, such as models with composite W and Z \cite{alternative}.  
Even if we take the point of view that SSB is correct, there is presently no deep underlying principle that can tell us whether the Higgs doublet of the SM is elementary or composite, notwithstanding the hierarchy problem which besets the elementary scalar. Is the Higgs field elementary or composite? That is the second question.  The third question which frequently comes up is the number of Higgs doublets or, for that matter, the number of Higgs fields whose vacuum expectation values (VEV) satisfy the constraint $M_W = M_Z \cos \theta_W$. 

For the past thirty years or so, a variety of very interesting models have been proposed to tackle the deep question of the nature of the Higgs boson if it exists, among which most notably are the ideas of Supersymmetry, Technicolor, Top-condensates, Little Higgs, etc..\cite{models}. Among these, there is a class of composite Higgs models, the so-called top condensate models, which is particularly attractive in that  the fermion which participates in the condensates is the top quark. In its early incarnation \cite{BHL, MTY}, the top quark was required to be heavier than $\sim$ 230 GeV in order for the scenario to work. The discovery of the top quark with a mass $\sim$ 171 GeV put a brake to this interesting model although further attempts have been made \cite{Kondo} to make the scenario work with a lower (experimental) mass of the top quark.

If the top quark is too light to form condensates that can break the electroweak symmetry, might it be possible that heavier fermions, if they exist, could fulfill this promise? For instance, a heavy fourth generation might be a possible candidate although other heavy fermions could very well play a similar role.  However, we will concentrate on a heavy fourth generation in this paper and show that its condensates can be formed and can spontaneously break the electroweak symmetry. 

The plan of the manuscript is as follows. To motivate the present work, we first summarize the results of \cite{pqchi1,pqchi2} where we showed that the Higgs-Yukawa sector of the SM with four generations exhibits at the two-loop level a quasi fixed point. This quasi fixed point is located in the TeV region if the fourth generation is sufficiently heavy i.e.  with a mass of around 400 GeV or above for the 4th quarks and is located at a much higher energy for a lighter fourth generation. Close to this quasi fixed point, the Yukawa couplings of the fourth generation increases rapidly leading to the possibility that fermion bound state formation can be generated by scalar exchanges. Next, we study the conditions under which scalar exchanges can lead to fermion condensation by using the Schwinger-Dyson (SD) equation.  This section is independent of the Renormalization Group (RG) results of \cite{pqchi1,pqchi2}. We find the critical Yukawa coupling above which condensation occurs. The dynamical mass of the fourth generation fermions (a boundary condition of the SD equation) is related to a cutoff scale $\Lambda$ and to the value of the coupling above the critical one. By requiring that this boundary condition (or 4th generation dynamical mass) to be 400-500 GeV, we find that $\Lambda$ depends on how close the Yukawa coupling is with respect to the critical one: the higher $\Lambda$ is the closer to the critical coupling the Yukawa coupling becomes. We then compare this crtical coupling with the results obtained for the Renormalization Group evolution (at two loops) of the Yukawa coupling of the fourth generation \cite{pqchi1} and show that although both ``light" and ``heavy" fourth generation Yukawa couplings reach this critical coupling, only the heavier one induces a TeV-scale condensate at the energy  where the corresponding quasi fixed point is situated. Finally, we will  comment on the possibility of the restoration of scale symmetry above the quasi fixed point. In particular, no condensation occurs above this quasi fixed point since it is only there that the boundary condition for the SD equation is satisfied.

\section{Quasi fixed point with a fourth generation}

For many years, the idea of the possible existence of a fourth generation of quarks and leptons was out of favor mainly because of electroweak precision data \cite{PDG} appeared to rule it out. However new analyses \cite{precision1} as well as the most recent global fits \cite{chanowitz,ehler} indicated that the SM with four generations- the so-called SM4- fits well with electroweak precision data. Lower bounds on the fourth generation quark ($t'$ and $b'$) masses were found to be  $338-385$ GeV \cite{cdf} (numbers obtained under a certain assumption about branching ratios).  (Strong dynamical effects of a heavy fourth generation on the $\rho$ parameter were investigated in \cite{hungrho}.) An extensive discussion of theoretical and experimental aspects of a fourth generation can be found in \cite{frampton1}. (A recent summary of the status of the fourth generation can be found in \cite{statement}.). An interesting question concerning the possibility of a long-lived fourth generation was discussed in \cite{frampton2} as well as its implications on gauge coupling unification \cite{hungunif}.

The resurgence of the idea of a fourth generation has generated a number of interesting works  on CP violation and rare B decays \cite{Bdecays}, Baryogenesis, electroweak symmetry breaking and phase transition \cite{baryo,ewsb}, and signals of composite Higgs scalars at the LHC \cite{soni}. It goes without saying that much remains to be done to fully understand the impact of the fourth generation during the Large Hadron Collider (LHC) era.

\begin{figure}
\includegraphics[angle=0,width=8.5cm]{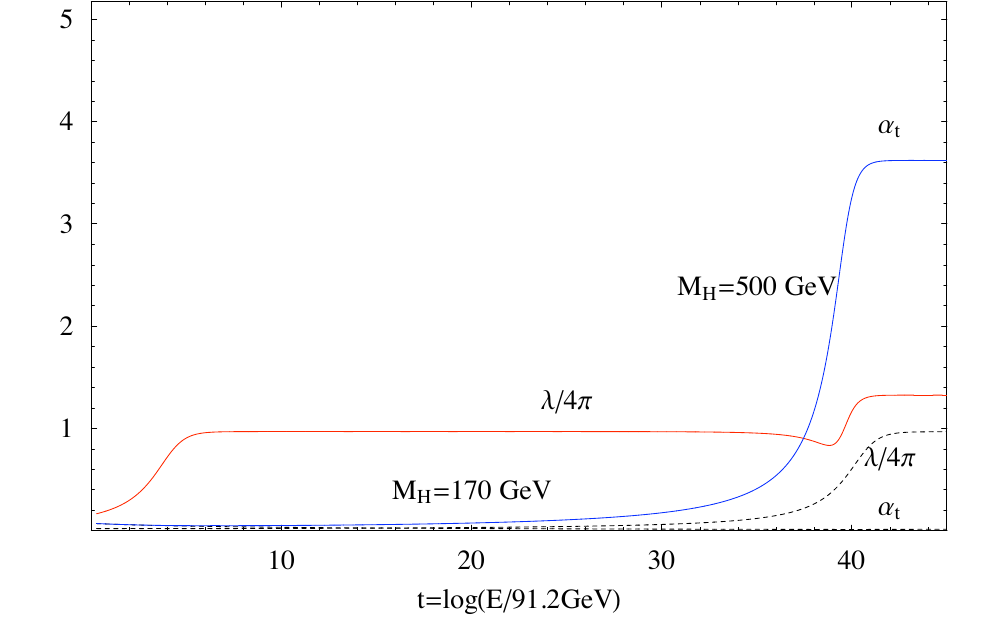}
\caption{The Higgs quartic and top Yukawa couplings at two loops as a function of energy for two
initial Higgs masses. Dashed line: 170 GeV; Solid line: 500 GeV}
\label{figtop}
\end{figure}

\begin{figure}
\includegraphics[angle=0,width=8.5cm]{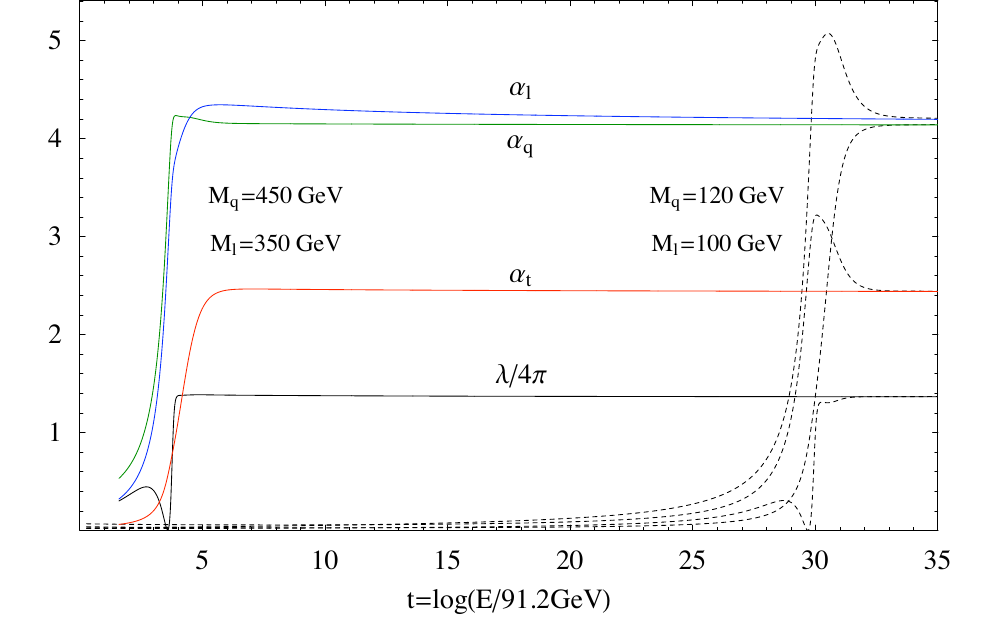}
\caption{The Higgs quartic and 4th generation and top Yukawa couplings at two loops as a function of energy for two
initial 4th generation masses}
\label{fig1}
\end{figure}

\begin{figure}
\includegraphics[angle=0,width=8.5cm]{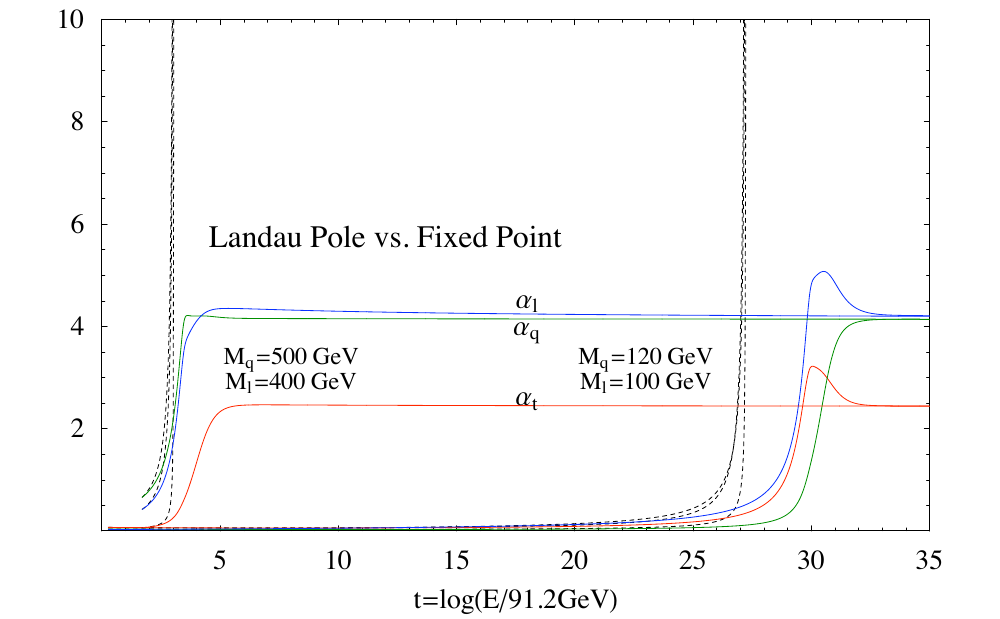}
\caption{{\small The Landau pole(dotted lines) and the quasi fixed point(solid lines) of the Yukawa couplings of the
fourth generation fermions and the top quark. For a heavy fourth generation (left side), both the Landau singularity from
one-loop RGEs  and the quasi fixed point from two-loop RGEs appear at about 2 $\sim$ 3 TeV, while for a light fourth 
generation (right side), their locations at the energy scale differ by two orders of magnitude.  
}}
\label{fig2}
\end{figure}

Since it is expected that the fourth generation, if it exists, should be heavy, it is natural to inquire about the behavior of the various couplings in SM4 in terms of energy. In particular, it is interesting to study the evolution of the Yukawa couplings of the fourth generation to see if they can become large enough at some energy scale to be able for 4th generation fermion-antifermion bound states to get formed. (It is well known that the top quark Yukawa coupling actually {\em remains}  ``small" at the Planck scale (Fig.\ref{figtop}).)  
This was what \cite{pqchi1} set out to do. The evolution of the couplings in SM4 was  carried out using the two-loop $\beta$ functions. The results are shown in Fig. \ref{fig1} for two different initial values of the 4th generation Yukawa couplings at the electroweak scale. One observes, at the two-loop level, the appearance of a quasi fixed point as shown in Fig. \ref{fig1}. 
This quasi fixed point occurs at a scale similar to the one where the one-loop Landau pole appears Fig. \ref{fig2}. This scale is denoted by $\Lambda_{FP}$ in \cite{pqchi1}.
 This will be identified as the physical cutoff scale in our discussion of the condensates using the SD equation.
 
 From Fig. \ref{fig1} and Fig. \ref{fig2} , one can see that the 4th generation Yukawa couplings start to become large as one approaches from below the cutoff scale $\Lambda_{FP}$. As we will show below, on their way to the quasi fixed point, these Yukawa couplings reach a critical value needed for the formation of condensates that break the electroweak symmetry.

 \section{The Schwinger-Dyson approach to condensate formation in the Higgs-Yukawa system}

 We investigate in this section the conditions under which a condensate of fermion-antifermion can be formed by the exchange of a scalar particle. In particular, the SM already possesses such a system:  An $SU(2)_L$ ``Higgs" doublet interacting with a left-handed fermion doublet and a fermion right-handed singlet.  We have also seen above that, for an initial value of the Yukawa coupling which is sufficiently large such as the one coming from the 4th generation (as one would have expected),  it steeply rises near the cutoff scale $\Lambda_{FP}$, leading to the expectation that condensates of 4th-generation fermions can get formed when the Yukawa coupling reaches a certain critical value. What this value is will be the subject of this section.
 
 As alluded to in the introduction, the two-loop RG evolution of the Yukawa coupling indicates the existence of a quasi fixed point at  $\Lambda_{FP}$. This gives rise to the possibility that the full $\beta$ functions vanish above that scale, albeit with couplings which may not  be at the values indicated in Fig. \ref{fig1}. In addition, if there are no explicit mass scales present at tree-level, one would have an interesting scale-invariant theory above $\Lambda_{FP}$. One could then imagine a scenario in which masses are generated by the spontaneous breaking of scale symmetry through the formation of fermion-antifermion condensates by the exchange of massless scalar particles. These condensates will then induce  below $\Lambda_{FP}$ a negative mass squared for the fundamental scalar which will itself acquire a non-zero vacuum expectation value (VEV). As we will describe below, in our model the so-called Higgs spectrum that can be observed below $\Lambda_{FP}$  will be composed of three $SU(2)_L$ doublets: one fundamental and two composites. This has been suggested in \cite{pqchi1,pqchi2} and the realization of such a scenario is now described in this paper.

 We now list the assumptions made in this section.

 1) Along the line suggested by the previous paragraph,  the fundamental ``Higgs doublet" will be assumed to be massless in our discussion of condensate formation. Because of this, the fundamental Higgs doublet will not develop a vacuum expectation value at tree level.
 
 2) The initial Yukawa couplings of the first three generations are small enough and will be below the critical value needed for condensate formation. In fact, a glance at Fig. \ref{fig1} tells us that this is the case even for the top quark. 
 
 3) All gauge interactions will be neglected. This is expected since the gauge couplings are very small compared with the Yukawa couplings near  $\Lambda_{FP}$.
 
 Under these assumptions, we start out with a scale-invariant SM with four generations. 

First, let us concentrate on the 4th generation quarks. The Yukawa Lagrangian is
\begin{equation} \label{action}
\mathcal{L}_Y = - g_{b'} ~\bar{q}_L \Phi ~b'_R - g_{t'} ~\bar{q}_L \widetilde{\Phi} ~t'_R + h.c. \,,
\end{equation}
where $\widetilde{\Phi}= i \tau_2 \Phi^{*}$ and $q_L =(t', b')_L$ as usually defined in the SM. Since $\Phi$ does not develop a VEV at tree level, this 4th generation (or any other generation for that matter) does not acquire a mass. Can the above Yukawa Lagrangian generate a dynamical mass for the 4th generation? To study this question, we proceed to compute the self-energy of the 4th generation fermions, $\Sigma_{4Q, 4L}(p)$, using (\ref{action}). As emphasized long ago by Pagels and Stokar \cite{pagels} for the case of chiral symmetry breaking in QCD, $\Sigma_{4Q,4L}(0)$  will set the scale for the dynamical electroweak symmetry breaking. In that paper, one notices the following steps. \cite{pagels} computed first $\Sigma(0)$ which was then used to calculate $f_{\pi}$. We will perform a similar type of calculations in this manuscript.

\begin{figure}
\includegraphics[angle=0,width=8.5cm]{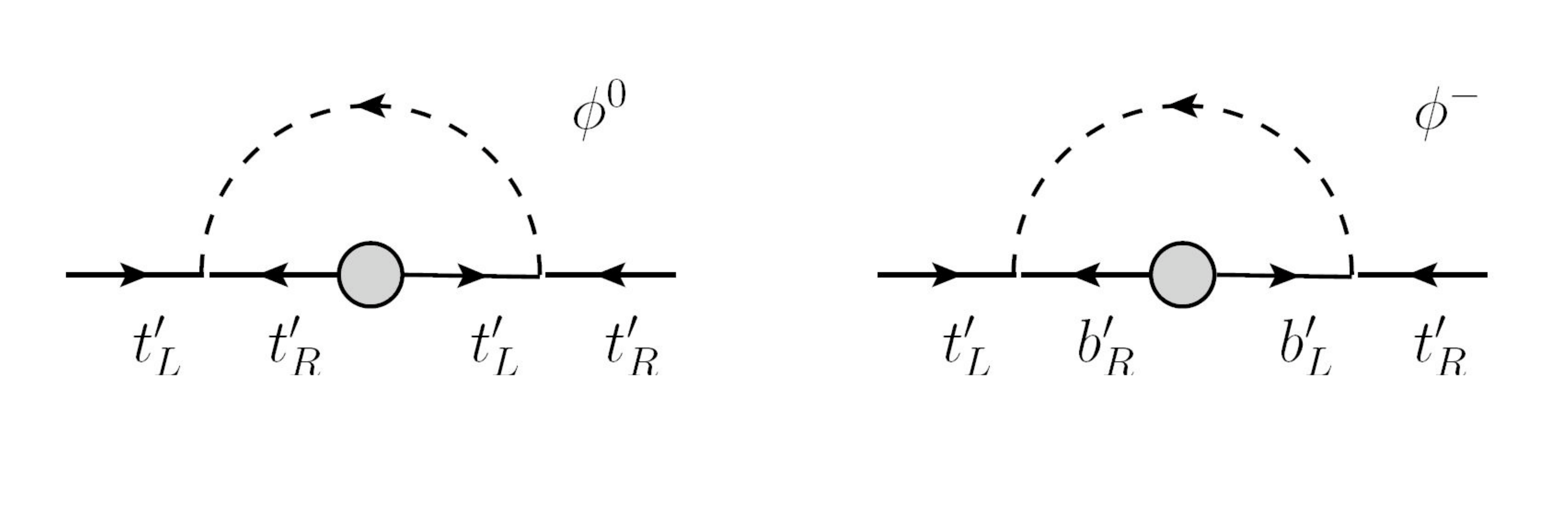}
\caption{{\small Graphs contributing to the right-hand side of the Schwinger-Dyson equation for the fermion self-energy $\Sigma(p)$ for $t'$. Similar graphs for the $b'$ self-energy are obtained by the substitution $t' \leftrightarrow b'$. The self-energies of the 4th leptons are computed in the same way.}}
\label{SDEfig}
\end{figure}

The Schwinger-Dyson (SD) equation will generically be of the form
\be \label{SD1}
\Sigma_{4Q}(p) = f(\Sigma_{4Q}(p) ) \,.
\ee
Fig.\ref{SDEfig} represents the right-hand side (i.e. $f(\Sigma_{4Q}(p) )$) of the SD equation, under the quenched 
(ladder or rainbow) approximation. In principle, one would have two self-energies $\Sigma_{t'}(p)$
and $\Sigma_{b'}(p)$. However, as in \cite{pqchi1, pqchi2}, we assume a degenerate 4th generation with $g_{b'} = g_{t'} = g_{4Q}$. This would imply the equality $\Sigma_{t'}(p)=\Sigma_{b'}(p)=\Sigma_{4Q}$. As we shall see, this leads to the desired $SU(2)$ custodial symmetry.

\begin{figure*}[!tb]
\centering
    \begin{tabular}{cc}
    \includegraphics[scale=0.75]{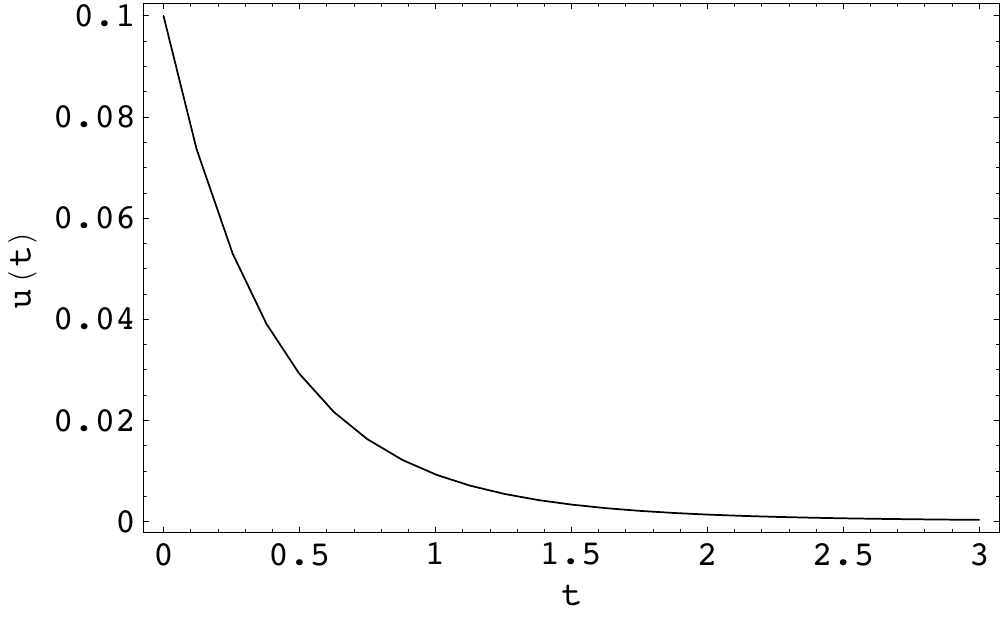} & \includegraphics[scale=0.75]{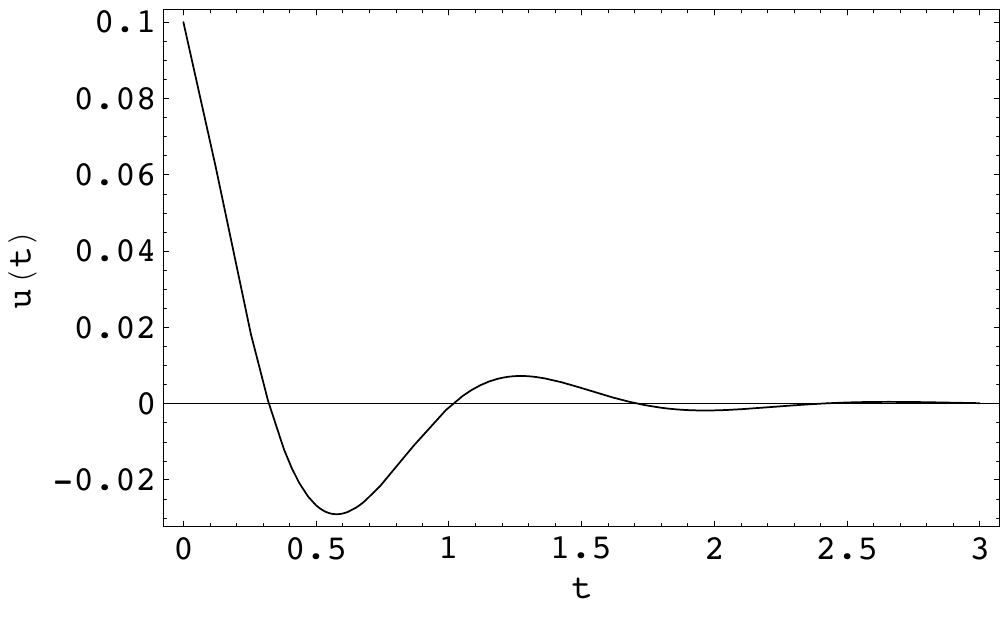}\\
    (a) & (b)
    \end{tabular}
\caption{{\small Solution of Schwinger-Dyson equation about self energy. (a) weak Yukawa coupling; (b) strong Yukawa coupling}}
\label{ut}
\end{figure*}

From the Schwinger-Dyson equation, we obtain
\be \label{SD}
\Sigma_{4Q}(p) = \frac{+2 g_{4Q}^2}{(2 \pi)^4}\int d^4 q \frac{1}{(p-q)^2} \frac{\Sigma_{4Q}(q)}{q^2 + \Sigma_{4Q}^2(q)} \,,
\ee
where $\Sigma_{4Q}(p)$ is the fermion self energy and the integration is from zero to an ultraviolet cutoff $\Lambda$. Notice that the factor of 2 on the right-hand-side of Eq. (\ref{SD}) comes from the equality of the two diagrams in Fig. (\ref{SDEfig}) due to the custodial $SU(2)$ assumption.
It can be converted to a differential equation, with $\alpha_{4Q}=g_{4Q}^2/4\pi$,
\be
\Box \Sigma_{4Q}(p) = - (\frac{\alpha_{4Q}}{\alpha_c}) \frac{\Sigma_{4Q}(q)}{q^2 + \Sigma_{4Q}^2(q)} \,,
\ee    
where the critical coupling $\alpha_c $ is given by
\be \label{crit}
\alpha_c = \pi/2 , 
\ee
with the following boundary conditions 
\bea \label{bc1}
\nn
&&\lim_{p \to 0} p^4 \frac{d \Sigma_{4Q}}{d p^2} = 0 \,, \\
&&\lim_{p \to \Lambda}  p^2 \frac{d \Sigma_{4Q}}{d p^2} + \Sigma_{4Q}(p)  = 0 \,.
\eea
This is the same as in \cite{Fukuda, Leung}, except that our critical value for the Yukawa couplings is 
$\alpha_c = \pi/2 $, while in the strong QED case \cite{Fukuda, Leung} it is $\alpha_c = \pi/3 $. Introducing a variable $t$ and a function $ u(t)$ by
\bea 
\nn
p &=& ~e^{t} \\
\Sigma_{4Q} (p) &=& e^t ~u(t-t_0)  \,,
\eea
one then obtains
\be \label{diff}
\frac{d^2 u}{d t^2} + 4 \frac{du}{dt} + 3 u + (\frac{\alpha_{4Q}}{\alpha_c}) \frac{u}{1+u^2} = 0 \,.
\ee
The boundary conditions become
\bea \label{bc2}
\nn
&&\lim_{t \to t_{\Lambda}} (u' + 3 u) = 0 \,, \\
&&\lim_{t \to -\infty} ( u' + u ) = 0 \,,
\eea
where $ t_{\Lambda} = \log \Lambda $.  Eq. (\ref{diff}) supplemented by the two boundary conditions (\ref{bc2}) is equivalent to the integral equation (\ref{SD}). Notice that the parameter $t_0=\ln \Sigma(0)$ (as in
\cite{Leung} (apart from the sign)) when the normalization is taken to be $e^t u(t) \rightarrow 1$
as $t \rightarrow -\infty$.\,
It is not difficult to see why $\alpha_{4Q} = \alpha_c$ is a critical point in the large
momentum region, where one can drop the $u^2$ term in the denominator of the last term of (\ref{diff}), since
u becomes vanishing as $ t \rightarrow \infty $. This leads to
\be 
\frac{d^2 u}{d t^2} + 4 \frac{du}{dt} + [3 + (\frac{\alpha_{4Q}}{\alpha_c})] u \approx 0, ~~\textrm{for large t} \,.
\ee
As in \cite{Leung} we then find two classes of asymptotic solutions at large momentum for different values of $\alpha$
\bea \label{solu} \nn
&& \Sigma_{4Q} (p) \sim p ^{- 1 + \sqrt{1- \frac{\alpha_{4Q}}{\alpha_c}}}, ~~~~~~~~~~~~~~~~~~~~~~~~\textrm{for}~\alpha_{4Q} \leq \alpha_c \,,, \\
&& \Sigma_{4Q} (p) \sim p ^{- 1} \sin [\sqrt{\frac{\alpha_{4Q}}{\alpha_c} -1} (\ln p + \delta)], ~~~\textrm{for}~\alpha_{4Q} > \alpha_c  \,, \nonumber \\
\eea
where $\delta $ is a constant phase (more on this below). We have also solved the differential equation (\ref{diff}) numerically and the
solutions $ u(t) $ are plotted in Fig.\ref{ut} for small and large Yukawa couplings (compared to $\alpha_c$) respectively.

For the strong Yukawa coupling solution to satisfy the boundary conditions (\ref{bc2}), we must have
\be \label{sigma0} 
\Sigma_{4Q} (0) = \Lambda e^{1 - n \pi/\sqrt{\frac{\alpha}{\alpha_c} -1} + \delta}~,~~~n=1, 2, ... 
\ee
It is easy to see that the solution with $n=1$ yields the largest fermion self energy and hence corresponds to the vacuum. One can rewrite Eq. (\ref{sigma0}) for $n=1$ as follows:
\be \label{phase}
\delta_0 = \ln (\frac{\Lambda}{\Sigma_{4Q} (0)}) + \frac{\pi}{\sqrt{\frac{\alpha}{\alpha_c} -1}} -1 \, ,
\ee  
where $\delta_0$ denotes the phase that corresponds to the vacuum solution.

The self energy $\Sigma_{4Q}(0)$ is the dynamical mass of the fourth generation in the sense that $m_{t'} = \Sigma_{4Q}(0)$ + Lagrangian mass.
For example, if $m_{t'} \sim 500$ GeV, then $\Sigma_{4Q}(0)< 500$ GeV.
We identify the cutoff scale $\Lambda$ with the aforementioned scale $\Lambda_{FP}$, which could be as low as the order of TeV. 

Here, the quantity which is most relevant to the dynamical breaking of the electroweak symmetry is the condensate of the fourth generation. It is given for the 4th generation quarks by
\bea \label{tbart1}
\nn
\langle \bar{t'}_L t'_R \rangle=\langle \bar{b'}_L b'_R \rangle  &=& - \frac{3}{4 \pi^4} \int d^4 q \frac{\Sigma_{4Q}(q)}{q^2 + \Sigma_{4Q}^2(q)} \\
                            &=& \frac{3}{2 \pi^2} (\frac{\alpha_c}{\alpha})~e^{3 t_{\Lambda}} [ u'(t_{\Lambda} -t_0) + u(t_{\Lambda} -t_0)] \nonumber \\
                            &=& \frac{-3}{\pi^2} (\frac{\alpha_c}{\alpha})~e^{3 t_{\Lambda}} u(t_{\Lambda} -t_0) \,,
 \eea
where we have used Eqs. (\ref{diff}) and (\ref{bc2}) and where the factor of 3 is the color factor. As the last line of Eq. (\ref{tbart1}) is evaluated at large $t$, 
using the asymptotic solution (\ref{solu}) one obtains
\bea \label{tbart2}
\nn
\langle \bar{t'}_L t'_R \rangle & \approx & - \frac{3}{\pi^2} (\frac{\alpha_c}{\alpha_{4Q}})~\Lambda~\Sigma_{4Q}^2(0)~ \times \nonumber \\
&&\sin[\sqrt{\frac{\alpha_{4Q}}{\alpha_c}-1} (t_\Lambda - t_0 + \delta_0)]  \nn \\ 
                            & \approx & - \frac{3}{\pi^2} (\frac{\alpha_c}{\alpha_{4Q}})~\Lambda~\Sigma_{4Q}^2(0)~ \times \nonumber \\
&&\sin[\sqrt{\frac{\alpha_{4Q}}{\alpha_c}-1} ( \ln \frac{\Lambda}{\Sigma_{4Q}(0)}+ \delta_0)]  \nonumber \\ 
                         &\approx&- \frac{3}{\pi^2} (\frac{\alpha_c}{\alpha_{4Q}})~\Lambda~\Sigma_{4Q}^2(0)~
\sin[\pi - \sqrt{\frac{\alpha_{4Q}}{\alpha_c}-1} ] \nonumber \\
                        &\approx&- \frac{3}{\pi^2} (\frac{\alpha_c}{\alpha_{4Q}})~\Lambda~\Sigma_{4Q}^2(0)~
\sin[\sqrt{\frac{\alpha_{4Q}}{\alpha_c}-1}]   \,, \nonumber \\
\eea
where we have used Eq. (\ref{phase}) in (\ref{tbart2}). 

Since $\langle \bar{t'}_L t'_R \rangle$ carries the electroweak quantum number ($SU(2)_L$ doublet), it will spontaneously break the Electroweak Symmetry. Therefore we impose the following condition
\be \label{condition}
\langle \bar{t'}_L t'_R \rangle \sim O(-\Lambda_{EW}^3)\,.
\ee

An expression for the 4th generation lepton condensates similar to Eq. (\ref{tbart1}, \ref{tbart2}) can be obtained written down without the color factor of 3 and with the replacements $\alpha_{4Q} \rightarrow \alpha_{4L}$ and $\Sigma_{4Q}(0) \rightarrow \Sigma_{4L}(0)$. We have
\be \label{LbarL}
\langle \bar{L}_L L_R \rangle=  \langle \bar{N}_L N_R \rangle \approx - \frac{1}{\pi^2} (\frac{\alpha_c}{\alpha_{4L}})~\Lambda~\Sigma_{4L}^2(0)~
\sin[\sqrt{\frac{\alpha_{4L}}{\alpha_c}-1}] \,.
\ee
The condition on the lepton condensate is similar to (\ref{condition}).

A few remarks are in order concerning Eq. (\ref{tbart2}) (or Eq. (\ref{LbarL}).  First, the condensate carries the electroweak quantum number similar to a Higgs doublet and contributes to the electroweak scale $\Lambda_{EW} \sim 246\, GeV$, with the condition (\ref{condition}) imposed on the condensate. Because of this, one can raise the question of fine tuning when the cut off scale $\Lambda$ becomes much larger than $O(TeV)$. In fact, when $\Lambda \sim O(TeV)$, one can see from Eq. (\ref{tbart2}) that $\alpha_{4Q}$ {\em does not have} to be close to $\alpha_c$ for the condensate to be of the order or less than the electroweak scale. There is {\em no} fine tuning in this case.  However, for $\Lambda \gg \Lambda_{EW}$, it is straightforward to see that the condensate will become large as well {\em unless}  $\sqrt{\frac{\alpha_{4Q}}{\alpha_c}-1}$ becomes small enough to balance $\Lambda$ . In general one requires that $\Lambda \sqrt{\frac{\alpha_{4Q}}{\alpha_c}-1} \sim \Lambda_{EW}$ or
\be \label{condition}
\frac{\alpha_{4Q}}{\alpha_c} \sim 1+ (\frac{\Lambda_{EW}}{\Lambda})^2 \,.
\ee
From (\ref{condition}), one can see that $\alpha_{4Q} \rightarrow \alpha_{c}$ as $\Lambda \rightarrow \infty$. For $\Lambda \sim 10^{16}\, GeV$, one gets $\alpha_{4Q}/\alpha_{c} \sim 1 + 10^{-28}$ which is a case of severe fine tuning! No such fine tuning is required if $\Lambda \sim O(TeV)$ as is the case with a heavy fourth generation. 

The next question concerns the possibility of an induced vacuum expectation value for the fundamental Higgs doublet. We started with a case in which scale invariance is assumed at tree level i.e. when the fundamental scalars are massless and there is no VEV. As we have seen above, the dynamical mass of the 4th generation quarks and leptons are $\Sigma_{4Q}(0)$ and $ \Sigma_{4L}(0)$ respectively. At energies below these dynamical masses, one obtains the following dimension 5 operators in the effective potential, involving the neutral fundamental scalar $\phi^0$
\be \label{dim5}
\frac{1}{2} \{\frac{g_{4Q}^2}{ \Sigma_{4Q}(0)} ( \bar{t'}_L t'_R +\bar{b'}_L b'_R) + \frac{g_{4L}^2}{ \Sigma_{4L}(0)} (\bar{L}_L L_R+ \bar{N}_L N_R\}|\phi^0|^2 \,,
\ee
which, upon using (\ref{tbart1}) and (\ref{LbarL}), gives rise to the following negative(sign of the condensates) effective mass squared term for the field $\phi^0$ 
\be \label{dim5vev}
\frac{1}{2} \{\frac{2 g_{4Q}^2}{ \Sigma_{4Q}(0)} \langle \bar{t'}_L t'_R \rangle + \frac{2 g_{4L}^2}{ \Sigma_{4L}(0)}\langle \bar{L}_L L_R \rangle \}|\phi^0|^2 \,.
\ee
Eq. (\ref{dim5vev}), when combined with the quartic term $\lambda \phi^\dagger \phi$, gives rise to a non-zero vacuum expectation value for the fundamental Higgs field $\phi$ which arises because of the presence of the fourth generation condensates. As expected, this vacuum expectation value vanishes when the condensates themselves vanish.

The above presentation using the SD approach deals mainly with the condition for condensate formation, namely the existence of a critical coupling $\alpha_c =\pi/2$. In principle, any value of $\alpha_{4Q,4L}$ above $\alpha_c$ will satisfy the condition for condensate formation. However, we have seen above that severe fine tuning (with $\alpha_{4Q,4L}$  required to be extremely close to $\alpha_c$ ) occurs when the cut off scale $\Lambda$ is very large. For $\Lambda \sim O(TeV)$, such severe fine tuning disappears. We have encountered a similar case of fine tuning in \cite{pqchi1, pqchi2}.  If $\Lambda_{FP}$, the scale where the quasi fixed point occurs, is very large, severe fine tuning is required for the scalar quartic coupling in order to insure vacuum stability. Only in the case of a heavy 4th generation with $\Lambda_{FP} \sim O(TeV)$ does one evade from such fine tuning. One can also see that close to $\Lambda_{FP}$,
the Yukawa couplings become large enough for the 4th generation 
fermions to form condensates. Above $\Lambda_{FP}$, the Yukawa couplings run into a quasi fixed point and
the scale symmetry is presumably restored (see next section for more discussions about scale invariance). We could identify the cut off scale used in the SD approach with $\Lambda_{FP}$, namely
\be \label{two}
\Lambda \sim \Lambda_{FP}, ~~\textrm{and above}~ \Lambda_{FP},  ~\alpha \rightarrow \alpha^*,~~\alpha^* \neq \alpha_c \,.
\ee
Note that our quasi fixed point ($\alpha^*)$ is {\it not} the critical point ($ \alpha_c $) as can be seen from Fig.\ref{ideal}. 

We now present a scenario of the aforementioned dynamical electroweak symmetry breaking in an``energy chronological" order: From high to low energy (or from high to low temperature).

\section*{A possible scenario}

\begin{figure}
\includegraphics[angle=0,width=8.5cm]{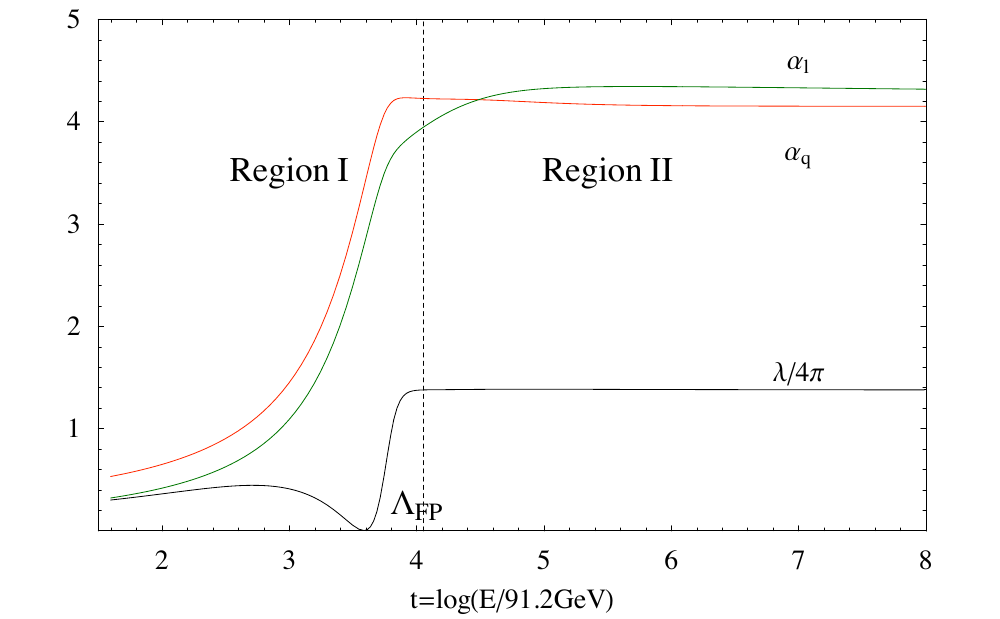}
\caption{{\small ($m_q=450$ GeV and $m_l=350$ GeV). The locations of $\Lambda_{FP}$, Region I and II;}}
\label{region}
\end{figure}
\begin{figure}
\includegraphics[angle=0,width=8.5cm]{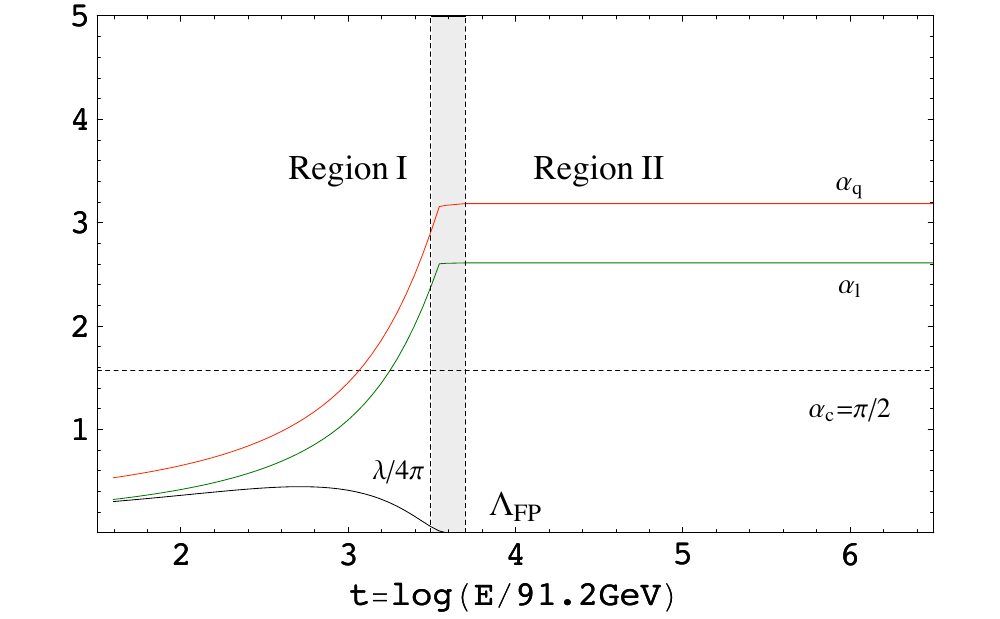}
\caption{A combination of RGE and Schwinger-Dyson analysis}
\label{ideal}
\end{figure}

We propose below a scenario which is an amalgam of the SD approach for dynamical electroweak symmetry breaking and the RG approach for a heavy fourth generation of \cite{pqchi1, pqchi2} . Let us first review the salient points presented in \cite{pqchi1, pqchi2} .

Our previous RG analysis \cite{pqchi1, pqchi2} reveals two interesting regions around $\Lambda_{FP}$ (see Fig.\ref{region}): 

Region I -- This is where the quartic coupling has a ``dip" just before the energy scale where it reaches 
its fixed point value. If the quartic coupling vanishes at the dip, it will give a Coulomb-like potential 
with strong coupling between the heavy fermions. According to the non-relativistic analysis done in\cite{pqchi1}  the 4th generation fermions can form condensates at the scale where the dip is located.

Region II -- This is the quasi fixed-point region. The Yukawa and the quartic couplings become practically
constant in this region and the corresponding $\beta$-functions vanish at the two-loop level. It is probable that a true fixed point, if it exists, corresponds to lower values of the couplings.

For the Yukawa couplings, one can readily see that two approaches (RG and SD) agree with each other if
we identify the SD cutoff and the Yukawa coupling as in (\ref{two})
\be \label{SDRG}
\Lambda^{SD} \approx \Lambda^{RG}_{FP}, ~\alpha^{SD} \approx \alpha^{RG*}, 
\ee
where we have labelled $\Lambda$ and $\alpha$ to distinguish the SD and RG cases. In the ensuing discussion, we will simply take $\Lambda_{FP}$ to be the physical cut off scale.

\begin{itemize}
\item Let us start with Region II. Let us assume that scale invariance- both at the classical and quantum levels- is established in this region. This is also the region where a true fixed point is presumably reached. (We have seen a hint -but not a proof- of that behaviour from the vanishing two-loop $\beta$ functions.) A mass scale can only arise spontaneously through the formation of a condensate.
Let us also assume that the now-constant Yukawa couplings (fixed points) $\alpha_{4Q,4L}$ have the values denoted by  $\alpha_{4Q,4L}^{CF}$ that can allow condensate formation and that those values are situated somewhere between $\alpha_c$ and the quasi fixed point. Above $\Lambda_{FP}$, {\em no} condensate can be formed despite the fact that the values of the Yukawa couplings are equal to $\alpha_{4Q,4L}^{CF}$. This is because the boundary conditions (\ref{bc1}) are only obeyed at $\Lambda^{SD} \approx \Lambda_{FP}$ and {\em not above } it.  In other words, there is no solution to the SD equation above $\Lambda_{FP}$ and hence no condensate.
\item Around $\Lambda_{FP}$, the condensates start to form and spontaneously break scale invariance. As a result, the Yukawa couplings start to run as the energy scale decreases as shown in Fig.\ref{ideal}.

The situation is more subtle for the quartic coupling, 
since in the SD approach the quartic term of the scalars can be dynamically generated at low energy, even when it is not
introduced by hand initially. We do not know other reasons for excluding the quartic coupling from the theory,
except that the self interactions of the scalar is not needed in the ladder approximation. In Fig.\ref{ideal} we
simply take the option $\lambda =0$ above $\Lambda_{FP}$. The shaded area reflects the uncertainty of the location
of $\Lambda_{FP}$ in the SD approach. In the RG approach \cite{pqchi1, pqchi2} we have emphasized that the location
of $\Lambda_{FP}$ and the values of the couplings at $\Lambda_{FP}$ might shift a bit if three or higher loop 
contributions are included.      

\end{itemize}
\section*{The Higgs sector}
The Higgs sector in our model is interesting on its own right and deserves an extensive separate investigation, both in terms of constraints and in terms of phenomenological implications. However, a few remarks are in order in this section concerning general features of the Higgs sector such as its spectrum and associated phenomena such as constraints on flavor changing neutral current (FCNC) interactions .

To study the spontaneous electroweak symmetry breaking from the Schwinger-Dyson approach, one has to 
identify the corresponding massless Nambu-Goldstone modes. In the top-quark condensation model \cite{BHL},
the physical Higgs scalar and Goldstone modes are found by considering the ``bubble sum" generated by
the four-fermion interaction. They appears as massive and massless poles in the scalar channel, the 
neutral-pseudoscalar channel and flavored channels respectively. 
As we have discussed above, the SM4 case yields three Higgs doublets at electroweak scale. 
They are denoted generically as
\bea \label{3higgs}
\nn
H_1 &=& ( \pi^+,\pi^-, \pi^0, \sigma ) \,, \\
\nn
H_2 &=& ( \bar{b'} t', \bar{t'} b', \bar{t'} t'-\bar{b'} b', \bar{t'} t'+ \bar{b'} b' ) \,, \\
H_3 &=& ( \bar{E } N, \bar{N} E, \bar{N} N -
\bar{E} E, \bar{N} N + \bar{E} E ) \,.
\eea
The usual Nambu-Goldstone (NG) bosons will be linear combinations of the pseudoscalar states ($ \pi^+$, etc..). The $3 \times 3$ mass matrices of the pseudoscalar bosons are denoted by  $\mathcal{M}^{+,-,0}$ for the charged and neutral states. In addition there will be a  $3 \times 3$ mass matrix for the $0^{++}$ states ($\sigma$, etc..) which is denoted by $\mathcal{M}_{S}^0$. The conditions for the existence of the three  NG bosons are
\be \label{goldstone}
\textrm{det} \mathcal{M}^{+,-,0} = 0 \,.
\ee
In fact (\ref{goldstone}) represents {\em three} constraint equations for the existence of three NG bosons. Out of the nine pseudoscalar states, there will remain six massive states. In addition, the diagonalization of $\mathcal{M}_{S}^0$ will give rise to three $0^{++}$``Higgs" scalars. 

In order to get a good handle on the mass spectrum of the Higgs sector, an effective potential for 3 Higgs doublets needs to be written down. At this stage, its form is dictated by the SM gauge invariance and looks similar to a generic 3-Higgs-doublet potential \cite{shizuya}. The spectrum is determined by the various coefficients which appear in the potential. The difference between a generic scenario and our approach is the fact that several of these coefficients can be computed and depend, at the one-loop level, principally on the fourth generation Yukawa couplings. They are basically "form factors" evaluated at zero momentum transfer. Conditions such as (\ref{goldstone}) constrain these form factors. In general, because of these mixings in the potential, it is reasonable to expect the scalar spectrum to contain particles which can be as light as it is experimentally allowed and as heavy as O(TeV). It is not possible to give a more precise statement at this stage of development. The phenomenology of this Higgs sector is quite interesting but is beyond the scope of the paper and will be dealt with elsewhere.

Last but not least, in a generic multi Higgs scenario, one usually expects flavor changing neutral current (FCNC) interactions which have to be brought under control. Unlike a generic elementary 3-Higgs model, the composite scalars discussed in this paper are bound states of the 4th generation quarks and leptons. As a result, the couplings of these composite scalars to the fermions of the "lighter" generations are quite constrained. 
This is coded in the mixings between the 4th generation and the lighter three in the Yukawa sector. In fact, there are many potential sources of FCNC suppression. The following simple example will illustrate the previous point. The most obvious process used to discuss FCNC is $K_L \rightarrow \mu e$. In the SM4 model discussed here, this process, $\bar{s} d \rightarrow \mu^+ e^-$, would have been mediated by a neutral scalar. For the components of this neutral scalar which come from the 4th generation bound sates, one can look at Eq. (\ref{3higgs}). The coupling of  $H_2 (b')$ to $\bar{s} d$ is proportional to the product of the mixing matrix elements that appear in the Yukawa couplings, namely $\propto U_{b's}^*\, U_{b'd}$. The coupling of $H_3 ( E)$ to $\mu^+ e^-$ is proportional to the product of the mixing matrix elements that appear in the Yukawa couplings, namely $\propto U_{E\mu}^*\, U_{E e}$. As a result, the matrix element of the process $K_L \rightarrow \mu e$ proceeding through the 4th generation composite scalars would be proportional to $U_{b's}^*\, U_{b'd}\, U_{E\mu}^*\, U_{E e}$ $\times$ ($H_2 (b')$ -$H_3 ( E)$ mixing). It is not hard to imagine that, if the mixing between the fourth generation with the first and second generations is {\em small}, that FCNC process can be highly suppressed. It is beyond the scope of this paper to go into the details of the Higgs sector but it suffices to say that FCNC processes can be naturally suppressed if the fourth generation mixes mostly with the third generation. This, in itself, has interesting consequences  concerning the search for the fourth generation.


Above  $\Lambda_{FP}$, we suspect the restoration of the scale symmetry from the vanishing $\beta$-functions
of the Yukawa couplings at two-loop level. Spontaneous breaking of scale invariance would suggest the existence
of a dilaton. In general it is difficult to construct a scale invariant theory at quantum level (see \cite{Shaposh} , \cite{kephart} for recent attempts), but our two-loop results hint that there might be a true fixed point or a scale invariant theory
above $\Lambda_{FP}$. Such scale invariant theory will be the ultraviolet completion of the SM and the 4th generation
and their condensates will be the messenger between this theory and the SM. 

\section{Epilog}

We have presented a model with four generations and with one fundamental massless Higgs doublet. 
If the Yukawa couplings are sufficiently large, one can form fermion-antifermion bound states and condensates . Using the Schwinger-Dyson equation in the rainbow approximation to compute the self-energy of the fourth generation fermions and subsequently their condensates, we found a critical Yukawa coupling $\alpha_c = g_{4Q,4L}^2/4\pi=\pi/2$ above which condensates carrying the electroweak quantum numbers can be formed which would spontaneously break the electroweak symmetry. The question of fine tuning is revisited showing how a large physical cut off scale is unnatural. We mirror this discussion with the one on the evolution of the Yukawa couplings of a heavy fourth generation \cite{pqchi1,pqchi2} where it was found that at a scale of $\Lambda_{FP} \sim  O(TeV)$ they become sufficiently large, exceeding $\alpha_c$, so as to satisfy the criterion for condensate formation. Above $\Lambda_{FP} $, there is a hint that the SM might be embedded into a scale-invariant theory. With the electroweak symmetry broken by condensates, one will have in our model a rich Higgs spectrum in which there are three doublets, two of which are composite and one being the original fundamental scalar doublet. This spectrum would be interesting phenomenologically and will be dealt with elsewhere. Also, the lighter three generations will obtain masses by coupling to the condensates and to the fundamental scalar doublet. It is beyond the scope of this paper to deal with this topic and it will be presented elsewhere. Needless to say, it is rather straightforward to mention it here that``light" fermion masses can be obtained without going outside the framework of SM4.

\section*{Acknowledgments}

This work is supported in parts by the US Department of Energy under grant 
No. DE-FG02-97ER41027. We would like to thank Sherwin Love for helpful discussions.


\begin{thebibliography}{99}
\bibitem{models}
The list of references related to these topics is extensive and it is impossible to put it in this note.
\end{thebibliography}
\end{document}